\documentclass[twocolumn, showpacs, pre]{revtex4-1}

\usepackage{amsmath}
\usepackage{graphicx}
\usepackage{subfigure}
\usepackage{color}

\begin{document}
\title{Strong field electrodynamics of a thin foil}
\author{Sergei V.~Bulanov}
\altaffiliation[Also at ]{Prokhorov Institute of General Physics, Russian Academy of Sciences, Moscow
119991, Russia and Moscow Institute of Physics and Technology, Dolgoprudny, Moscow region 141700, Russia}
\affiliation{Kansai Photon Science Institute, JAEA, Kizugawa, Kyoto 619-0215, Japan}
\author{Timur~Zh.~Esirkepov}
\affiliation{Kansai Photon Science Institute, JAEA, Kizugawa, Kyoto 619-0215, Japan}
\author{Masaki~Kando}
\affiliation{Kansai Photon Science Institute, JAEA, Kizugawa, Kyoto 619-0215, Japan}
\author{Stepan S. Bulanov}
\affiliation{University of California, Berkeley, California 94720, USA}
\author{Sergey G.~Rykovanov}
\affiliation{Lawrence Berkeley National Laboratory, Berkeley, California 94720, USA}
\author{Francesco~Pegoraro}
\affiliation{Physics Department, University of Pisa, Pisa 56127, Italy}

\date{\today}

\begin{abstract}
Exact solutions describing the  nonlinear electrodynamics of a thin double layer foil are presented. 
These solutions correspond to  a   broad range of problems of  interest for the  interaction 
 of high intensity  laser 
 pulses with   overdense plasmas  such as   frequency upshifting, high order harmonic generation and 
high energy ion acceleration. 
\end{abstract}


\maketitle

\section{Introduction}

High power laser irradiation of various targets, such as  solid, 
cluster or gas targets, has been used for a number of years 
in order to  study a broad range of mechanisms of high energy ion and electron acceleration 
\cite{ion-rev, ele-rev}, high and low energy
photon generation \cite{photon-rev1, photon-rev2, photon-rev3}, and to explore  problems of interest 
for modeling processes relevant to fundamental physics \cite{MTB, fund-rev}  and astrophysics \cite{astro-rev}.

When a high-intensity laser pulse interacts with a very thin
foil target, which can be modelled as a thin slab of overdense plasma, 
features appear that are not encountered either in underdense
or in overdense plasmas as noted in the current literature,
see  e.g., Refs. \cite{Foil1,MTB}. These features provide novel regimes 
for ion acceleration \cite{IONS, IONS-UL, AAMS, MacchiPRLReflectivity, BulanovOptimalShape}, 
 relativistic high order harmonics generation \cite{RelOscMirr, Pirozhkov2006, MikhPRL}, 
 light frequency upshifting 
\cite{BulanovRMP, Bulanov2006, Kulagin2007, Kulagin2013, Kando2007, Kagami, BrII2012},
and  laser pulse shaping \cite{Foil1, Shaping, Reed2009, Foil3, transparency, Hur2012}. They become important when
the foil thickness is shorter than, or of the order of, both the
laser wavelength and the plasma collisionless skin depth.

The thin foil model developed in Refs. \cite{Foil1, Pirozhkov2006, Kulagin2013, Kando2007, Bulanov1975, Bratman1995}
has  the advantage of being an  exactly solvable nonlinear boundary 
problem in electrodynamics describing  the effects of a strong 
radiation friction force (see \cite{Foil1, LADvsLL}). 
 
In this paper we present a set of exactly solvable equations 
describing the nonlinear electrodynamics of a thin double layer foil when the effects of 
the charge separation electric field and of  the radiation back reaction are taken into account.
Within the framework of the thin foil approximation we shall address the generation of  high order 
 harmonics, when the thin foil models a
relativistic oscillating mirror \cite{RelOscMirr}, the frequency upshifting during 
the head-on collision of an  electromagnetic wave 
with a relativistic foil, corresponding to the case of a relativistic flying mirror \cite{BulanovRMP}, 
and the ion acceleration when the  radiation pressure  
of the electromagnetic wave pushes the electron layer 
pulling forwards the ions according to the radiation pressure acceleration regime \cite{IONS}.

\section{Equations of 1D Electrodynamics}

Let us consider a one-dimensional model of the interaction of a  laser  pulse with thin foil targets. 
Each foil comprises two layers: an ion layer with positive electric charge $ e n_{0} l_{0}$  
and a negatively charged, $-e n_{0} l_{0}$, electron layer,
 $l_{0}$ is the thickness of the foil  which has equal ion and electron density. 
Here and below for the sake of brevity we assume that ions and electrons have equal electric charge and that  the layer
 thickness and density are the same for all layers.

It is convenient to describe the thin foil distribution function as a delta-function 
in both momentum and coordinate. 
Below we use dimensionless variables with time and space  
normalized on $\omega_0^{-1}$ and $c/\omega_0$ respectively, the density unit is $n_{cr}
$, 
and the electromagnetic (EM) field is normalized on $m_e\omega_0 c/e$. The particle velocity and momentum are 
normalized on $c$ and $m_{\alpha}c$ where $\alpha$ denotes the species in the $\alpha_{\rm th}$ layer. 
Here $n_{cr}=m_e \omega_0^2/4 \pi e^2$  is the critical density for an   
EM wave with  frequency $\omega_0$. 
In these expressions 
$c$ is the speed of light in vacuum, $e$ and $m_e$ are the electron charge and mass, respectively.

Then the only parameter 
describing the  electrodynamic  properties  of the  $\alpha_{\rm th}$ layer 
will  be the normalized areal charge density density 
 $\epsilon_{\alpha}$, which expressed in terms of  the dimensional  layer density and thickness 
is given by (see Ref. \cite{Foil1})
\begin{equation}
\label{epsilone}
\epsilon_{\alpha}=\frac{2\pi n_{0} e^2 l_0}{m_{\alpha} \omega_0 c } .
\end{equation}
The electromagnetic field obeys the Maxwell equations, 
\begin{equation}
\label{mxwl}
\partial_{x_{\mu}}\partial_{x^{\mu}}A^{\nu}=\frac{4\pi}{c} j^{\nu}
\end{equation}
with the 
four-vector of the electric current density equal to 
\begin{equation}
\label{jnu_equation}
j^{\nu}=\sum_{\alpha}j^{\nu}_{\alpha},
\end{equation}
and $\nu=0,1,2,3$.
Here the electric current carried by the $\alpha_{\rm th}$ layer is given by
\begin{equation}
\label{jnua_equation}
j^{\nu}_{\alpha}=Z_\alpha(c,{\bf v}_{\alpha})\, e n_0 l_0 \delta(x-x_{\alpha}(t)),
\end{equation}
where $\delta(x)$ is the Dirac delta function and $Z_{\alpha}=\pm 1$. The $\alpha_{\rm th}$ layer velocity is 
${\bf v}_{\alpha}=v_{1,\alpha}{\bf e}_1+v_{2,\alpha}{\bf e}_2+v_{3,\alpha}{\bf e}_3$, and ${\bf e}_1, \, {\bf e}_2,\, {\bf e}_3$ are unit vectors 
in the $x$, $y$ and $z$ directions, $x_{\alpha}(t)$ is the $\alpha_{\rm th}$ layer coordinate.

Using the results of Refs. \cite{Foil1, Feynman1966, Bulanov1975, Bratman1995}
we can write 
the solution to the wave equation which yields
\begin{equation}
\label{E_equation}
{\bf E}_{\alpha}(x,t)
=Z_{\alpha}\epsilon_{\alpha} \left[  s(x,  \bar t_\alpha)
{\bf e}_1+\frac{v_{2,\alpha}(\bar t_{\alpha}){\bf e}_2+v_{3,\alpha}(\bar t_{\alpha}){\bf e}_3}{1-s_{\alpha}(x,\bar t_{\alpha})v_{1,\alpha}(\bar t_{\alpha})}\right],
\end{equation}
\begin{equation}
\label{B_equation}
{\bf B}_{\alpha}(x,t)=
-Z_{\alpha} \epsilon_{\alpha} s_{\alpha}(x,\bar t_{\alpha})\frac{v_{3,\alpha}(\bar t_{\alpha}){\bf e}_2
-v_{2,\alpha}(\bar t_{\alpha}){\bf e}_3}{1-s_{\alpha}(x,\bar t_{\alpha})v_{1,\alpha}(\bar t_{\alpha})},
\end{equation}
for the electric and magnetic field formed by a single $\alpha_{\rm th}$ layer,
where  $s_{\alpha}(x,\bar t_{\alpha})={\rm sgn}(x-x_{\alpha}(\bar{ t_{\alpha}}))$ 
with the signum function ${\rm sgn}(x)=+1$ for $x>0$ and ${\rm sgn}(x)=-1$ if $x<0$. 
For given longitudinal $v_{1,\alpha}$ and transverse 
$v_{2,\alpha}{\bf e}_2+v_{3,\alpha}{\bf e}_3$  components of the particle velocity, these expressions 
 describe the EM wave emitted by the  thin layer, 
which acts as  a 1D electric charge.
Here and below the retarded time is determined by the  equation 
\begin{equation}
\label{t_ret}
\bar t_{\alpha}=t-|x-x_{\alpha}(\bar t_{\alpha})|. 
\end{equation}
These relationships can also be easily derived with the Li\'enard-Wiechert potentials  ~\cite{LandauLifshitzVol2} 
for the 1D four-vector of the electric current density.

Taking into account that  the transverse components of the  fields $\,  {\bf E}_{{\alpha},l}$ and ${\, \bf B}_{{\alpha},l}$  
at the $\alpha_{\rm th}$ layer, $x=x_{\alpha}(t)$ and $\bar t_{\alpha}=t$, 
are equal to the average of their values  at both sides, 
%
\begin{equation}
\label{E_atlayer}
{\bf E}_{{\alpha},l}=Z_{\alpha}\epsilon_{\alpha} \frac{v_{2,{\alpha}}{\bf e}_2+v_{3,{\alpha}}{\bf e}_3}{1-v_{1,{\alpha}}^2},
\end{equation}
\begin{equation}
\label{B_atlayer}
{\bf B}_{{\alpha},l}=-Z_{\alpha}\epsilon_{\alpha} v_{1,{\alpha}}\frac{v_{3,{\alpha}}{\bf e}_2-v_{2,{\alpha}}{\bf e}_3}{1-v_{1,{\alpha}}^2},
\end{equation}
we can write the
expression for the EM acting on the $\alpha_{\rm th}$ layer as the sum of the external 
and self-action fields: ${\bf E}+{\bf E}_{{\alpha},l}$ and ${\bf B}+{\bf B}_{{\alpha},l}$,
where 
\begin{equation}
\label{E_ext}
{\bf E}=-\partial_t{\bf A}_{0,\perp}(x_{\alpha},t)+\sum_{\alpha^{\prime}\ne \alpha}
{\bf E}_{\alpha^{\prime}}(x_{\alpha},\bar t_{\alpha,\alpha^{\prime}}(t))
\end{equation}
and
\begin{equation}
\label{B_ext}
{\bf B}={\bf e}_x\times\partial_x{\bf A}_{0,\perp}(x_{\alpha},t)
+\sum_{\alpha^{\prime}\ne \alpha}{\bf B}_{\alpha^{\prime}}(x_{\alpha},
\bar t_{\alpha,\alpha^{\prime}}(t)).
\end{equation}
Here $\bar t_{\alpha,\alpha^{\prime}}(t)$ should be found from equation
\begin{equation}
\label{t_retret}
\bar t_{\alpha,\alpha^{\prime}}(t)=t-|x_{\alpha}(t)-x_{\alpha^{\prime}}(\bar t_{\alpha,\alpha^{\prime}})|. 
\end{equation}
The vector potential ${\bf A}_{0,\perp}$,  normalized on $m_e c^2/e$, corresponds to the external EM field. 
In particular it describes  the EM pulse incident on the target. In these expressions $\partial_x$ and $\partial_t $ 
denote partial derivatives with respect to the coordinate $x$  and  the time $t$.

Using the above obtained relationships we can write the  equations of  the $\alpha_{\rm th}$ layer motion in components as
\[\dot p_{1,\alpha}=
Z_{\alpha}\mu_{\alpha}\left({E}_{1}+
\frac{p_{2,\alpha} B_3-p_{3,\alpha} B_2}{\gamma_{\alpha}}\right)\]
\begin{equation}
\label{p1_equation}
\qquad-\epsilon_{\alpha} 
\frac{p_{1,\alpha} (p_{2,\alpha}^2+p_{3,\alpha}^2)}{\gamma_{\alpha} (\gamma_{\alpha}^2-p_{1,\alpha}^2)},
\end{equation}
\begin{equation}
\label{p2_equation}
\dot { p}_{2,\alpha}=Z_{\alpha}\mu_{\alpha} \left({E}_{2}-\frac{p_{1,\alpha}B_3}{\gamma_{\alpha}}\right)
-\epsilon_{\alpha} \frac{{ p}_{2,\alpha}}{\gamma_{\alpha}},
\end{equation}
\begin{equation}
\label{p3_equation}
\dot { p}_{3,\alpha}=Z_{\alpha}\mu_{\alpha} \left({E}_{3}+\frac{p_{1,\alpha}B_2}{\gamma_{\alpha}}\right)
-\epsilon_{\alpha} \frac{{ p}_{3,\alpha}}{\gamma_{\alpha}}.
\end{equation}
Here, $\mu_{\alpha} = m_e/m_{\alpha}$, a dot, $\dot {}$\, ,  denotes time derivative, 
$p_{1,\alpha}$ and ${p}_{2,\alpha}{\bf e}_2+{p}_{3,\alpha}{\bf e}_3$ are the longitudinal and perpendicular momenta of the  
of the particles in the $\alpha_{\rm th}$ layer. 
The layer coordinate $x_{\alpha}(t)$ depends on time according 
to equation $\dot x_{\alpha}= p_{1,\alpha}/\gamma_{\alpha}$
where $\gamma_{\alpha}=\sqrt{1+p_{1,\alpha}^2+p_{2,\alpha}^2+p_{3,\alpha}^2}$ 
is the relativistic Lorentz factor. 
The longitudinal and perpendicular components of the electric field are equal 
to ${E}_{1}={\bf e}_1 ({\bf e}_1 \cdot {\bf E})$ and 
${\bf E}-{\bf E}_{1}$, respectively.
The last terms  on the r.h.s.  of 
Eqs. (\ref{p1_equation}), (\ref{p2_equation}) and (\ref{p3_equation})
are the longitudinal and perpendicular components of the 1D electrodynamics radiation friction force, respectively. 

Multiplying Eqs. (\ref{p1_equation} -- \ref{p3_equation}) by ${\bf v}_{\alpha}$ 
and adding them, we obtain the equation
\begin{equation}
\frac{d {\cal E}_{\alpha}}{d t}=Z_{\alpha}{\bf E} 
\cdot {\bf v}_{\alpha} -\epsilon_{\alpha}\frac{p_{2,\alpha}^2+p_{3,\alpha}^2}{1+p_{2,\alpha}^2+p_{3,\alpha}^2},
\label{eq:h}
\end{equation}
where ${\cal E}_{\alpha}$ is a kinetic energy of the $\alpha_{th}$  layer.
As we see the rate of radiative energy  losses depends only on the momentum component along the layer. 
The rate of energy   loss vanishes at 
$p_{2,\alpha}^2+p_{3,\alpha}^2=0$ and it is limited by the value of $\epsilon_{\alpha}$, 
because the layer electric field cannot exceed 
$2 \pi e n l$ (in dimensional units).
We shall return to this issue below.

In the above formulated 1D electrodynamics the EM wave is normally incident on the target. 
However, as is well known, by choosing 
proper initial conditions for the transverse component of the layer momentum, ${p}_{2,\alpha}$ and ${p}_{3,\alpha}$ 
in Eqs. (\ref{p1_equation}), (\ref{p2_equation}) and (\ref{p3_equation}),
we obtain a solution for an obliquely incident wave in the boosted frame of reference 
(see Refs. \cite{Foil1, Bourdier, PGARB, RelOscMirr}),
provided  initially all the sheets  are at rest and stationary and the (two) pulses are in vacuum (outside the foils).

This 1D electrodynamics system of equations for the EM field and layer motion 
can also be considered as an extension of Dawson's electrostatic 1D plasma model \cite{Dawson62} 
to the electromagnetic case with  
self-action (radiation reaction) taken into account. We notice here that in the case  of a rotating electric field 
Eqs. (\ref{p2_equation}) and (\ref{p3_equation})
are reduced to the equations analysed in Ref. \cite{LADvsLL}.

For analytical considerations and numerical integration 
of Eqs. (\ref{p1_equation} -- \ref{p3_equation}) it is convenient 
to take the vector potential ${\bf A}_{0,\perp}$ to 
propagate in the positive $x$  direction i.e., to depend on $t-x$ and to introduce  the function 
\begin{equation}
h_{\alpha}=\gamma_{\alpha}-p_{1,\alpha}
\label{eq:h}
\end{equation}
and the variable
\begin{equation}
\tau_{\alpha}=t-x_{1,\alpha}.
\label{eq:h}
\end{equation}
since, in the limit of week radiation friction $\epsilon_{\alpha} \to 0$  
and vanishing  longitudinal electric field $E_1$, the function $h_{\alpha}$ is an integral of motion.
Using these variables we can present Eqs. (\ref{p1_equation} -- \ref{p3_equation}) in the  implicit  form
\begin{equation}
\label{h_equation-tau}
\frac{d h_{\alpha}}{d \tau_{\alpha}}=- Z_{\alpha}\mu_{\alpha}E_1-\epsilon_{\alpha} 
\frac{{ p}_{2,\alpha}^2+{ p}_{3,\alpha}^2}{1+{ p}_{2,\alpha}^2+{ p}_{3,\alpha}^2}
\end{equation}
\begin{equation}
\label{x1_equation-tau}
\frac{d x_{1,\alpha}}{d \tau_{\alpha}}=\frac{1+{ p}_{2,\alpha}^2+{ p}_{3,\alpha}^2-h_{\alpha}^2}{2 h_{\alpha}^2},
\end{equation}
\begin{equation}
\label{x2_equation-tau}
\frac{d x_{2,\alpha}}{d \tau_{\alpha}}=\frac{p_{2,\alpha}}{h_{\alpha}},
\end{equation}
\begin{equation}
\label{x3_equation-tau}
\frac{d x_{3,\alpha}}{d \tau_{\alpha}}=\frac{p_{3,\alpha}}{h_{\alpha}},
\end{equation}
\begin{equation}
\label{tau_equation-tau}
\frac{d \tau_{\alpha}}{d t}=\frac{2 h_{\alpha}^2}{1+{ p}_{2,\alpha}^2+{ p}_{3,\alpha}^2+h_{\alpha}^2},
\end{equation}
with
\begin{equation}
\label{p1_equation-tau}
p_{1,\alpha}=\frac{1+p_{2,\alpha}^2+p_{3,\alpha}^2-h_{\alpha}^2}{2 h_{\alpha}}
\end{equation}
\begin{equation}
\label{p2_equation-tau}
p_{2,\alpha}=a_{2,\alpha}-\epsilon_{\alpha} \left( x_{2,\alpha} -  
\left. x_{2,\alpha} \right|_{\tau_{\alpha}=-x_{1,\alpha}} \right),
\end{equation}
\begin{equation}
\label{p3_equation-tau}
p_{3,\alpha}=a_{3,\alpha}-\epsilon_{\alpha} \left( x_{3,\alpha} - 
\left. x_{3,\alpha} \right|_{ \tau_{\alpha}=-x_{1,\alpha}} \right),
\end{equation}
and
\begin{equation}
\label{gamma_equation-tau}
\gamma_{\alpha}=\frac{1+p_{2,\alpha}^2+p_{3,\alpha}^2+h_{\alpha}^2}{2 h_{\alpha}}.
\end{equation}

\section{Single electron layer acceleration by the laser light pressure}

\subsection{Limit of week radiation friction}

In order to elucidate the  basic properties of the 1D  electrodynamics  formulated  above  we consider 
the motion of a single electron layer in the  plane EM wave ${\bf a}_{0,\perp}(t-x)$.
In this case the longitudinal component of the electric field, ${\bf E}_{1}$, 
in the r.h.s. of Eq. (\ref{p1_equation}) vanishes, 
and the electric and magnetic fields are equal to
${\bf E}=-\partial_t{\bf a}_{0,\perp}(t-x)$ and ${\bf B}={\bf e}_1\times\partial_{x}{\bf a}_{0,\perp}(t-x)$, 
respectively,
with given ${\bf a}_{0,\perp}$. The electric and magnetic fields   are taken at $x=x_{\alpha}$. 

In the case without radiation losses, when $\epsilon_{\alpha}=0$, Eqs. (\ref{eq:h} -- \ref{gamma_equation-tau}) 
yield the well known results \cite{LLCTF}, 
\begin{equation}
\label{eq:h}
h_{\alpha}={\rm constant}, \quad p_{2,\alpha}=a_{0,2} (\tau_\alpha), \quad p_{3,\alpha}=a_{0,3}(\tau_\alpha).
\end{equation}
 If the layer before interacting with the EM pulse is at rest  $h_{\alpha}=1$.  Then 
for $p_{1,\alpha}$, $\gamma_{\alpha}$, $x_{1,\alpha}$ and $\tau_{\alpha}$ we have 
\begin{equation}
\label{LW1}
p_{1,\alpha}=\frac{1}{2}( a_{0,2}^2(\tau_\alpha) +a_{0,3}^2(\tau_\alpha)),
\end{equation}
\begin{equation} 
\label{LW2}
\gamma_{\alpha}=1+p_{1,\alpha}=1+\frac{1}{2}( a_{0,2}^2(\tau_\alpha) +a_{0,3}^2(\tau_\alpha)),
\end{equation}
\begin{equation} 
\label{LW3}
x_{1,\alpha}(\tau_{\alpha})=\frac{1}{2}\int_{-\infty}^{\tau_{\alpha}}d\tau' (a_{0,2}^2(\tau')+a_{0,3}^2(\tau')), 
\end{equation}
and 
\begin{equation} 
\label{LW4}
t=\tau_{\alpha}+x_{1,\alpha}(\tau_{\alpha}).
\end{equation}

As a result of the interaction of the  electron layer  with a finite duration electromagnetic pulse, 
its kinetic energy, 
${\cal E}_{kin,\alpha}=\gamma_{\alpha}-1$, increases from zero to a maximum value equal to $a_m^2/2$ 
and then decreases to almost zero (an exponentially small value for a pulse longer than its wavelength) 
after the electromagnetic pulse has overtaken the layer. Here $a_m$ is the maximum amplitude of the pulse. 
This fact is referred to as the Lawson -- Woodward theorem \cite{Lawson, Woodward}. The layer displacement 
from the initial position is equal to 
\begin{equation} 
\xi_{1,\alpha}=\frac{1}{2}\int_{-\infty}^{+\infty}d\tau' (a_{2,\alpha}^2(\tau')+a_{3,\alpha}^2(\tau')).
\label{displace}
\end{equation}

In the limit of small but finite radiation losses we can find the radiation scattered by the layer.  
Considering $\epsilon_{\alpha}$ as the  parameter of  a perturbation expansion, 
we calculate the reflected and transmitted waves by using 
Eqs. (\ref{E_equation}), (\ref{B_equation}) and (\ref{t_ret}), 
in which the layer velocity components and $\bar t_{\alpha}$ are obtained from Eqs. (\ref{eq:h} -- \ref{LW4})
for a  pulse linearly polarized along the 2-direction.
This yields for the electric field of the wave scattered in forward direction 
\begin{equation}
{E}_{2,\alpha}(x,t)
=\left.-\epsilon_{\alpha}  a_2 \sin(\tau_{\alpha})\right|_{\tau_{\alpha}=t-x},
\label{E-forw-e0}
\end{equation}
i.e. the transmitted wave is $(1-\epsilon_{\alpha})  a_2 \sin(\tau_{\alpha})$.
The backward scattered wave, which is the wave reflected from the receding layer, is given by
\begin{equation}
{E}_{2,\alpha}(x,t)
=\left.-\frac{ \epsilon_{\alpha} a_2 \sin(\tau_{\alpha})}{1+a_2^2 \sin^2(\tau_{\alpha})}\right|_{\tau_{\alpha}+\frac{a_2^2}{2}(\tau_\alpha-\frac{\sin 2\tau_{\alpha}}{2})=t+ x}.
\label{E-backw-e0}
\end{equation}

\begin{figure}[tbph]
\centering
\includegraphics[width=6cm,height=9cm]{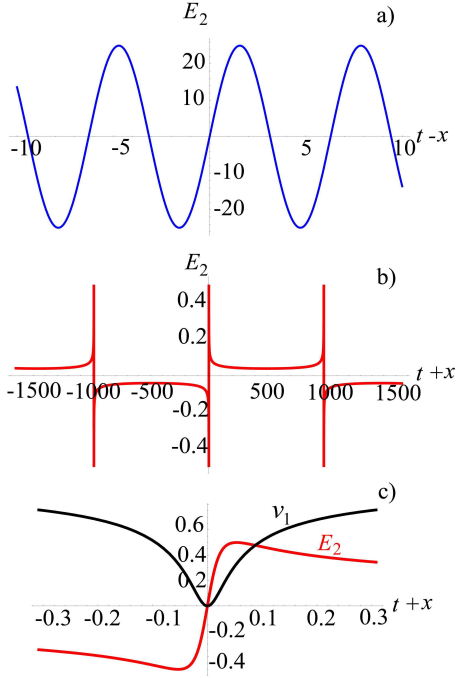} 
\caption{(Color online) Emitted in the forward  direction (a) and backward reflected (b) waves for $a_2=25$. 
The electric field amplitude is divided by $\epsilon_{\alpha}$. 
(c) Local structure of the electric field $E_2$ and of the longitudinal velocity of the electron layer $v_1$.}
\label{FIG1}
\end{figure}

Here for the sake of brevity we consider the interaction of the layer with a sinusoidal electromagnetic 
wave given by ${\bf a}_{\perp}=a_2 \sin(t-x){\bf e}_2$
for $t>0$ and zero before. Fig. \ref{FIG1} shows the waves emitted in the forward 
and backward directions, respectively.

Due to the double Doppler effect the wavelength  of the wave reflected back by  the receding 
layer (Fig. \ref{FIG1} b) is larger than the incident wavelength. In addition, the reflected wave 
is not sinusoidal. The minimal electric field where $\partial  {E}_{2,\alpha}(x,t)/\partial x= 0$ 
is equal to $\epsilon_{\alpha}/a_2$ in the limit $a_2 \gg 1$.
In this limit  every each half-period the wave profile becomes singular at the point 
where $\partial x /\partial t _{\alpha}(x,t)= 0$. 
In the vicinity of the singular point $t+x$ and $E_2$ 
depend on $\tau_{\alpha}$ as $t+x\approx \tau_{\alpha}-a_2^2 \tau_{\alpha}^3/3$ and 
$E_2\approx a_2 \tau_{\alpha}-a_2^2 \tau_{\alpha}^3$, which gives 
\begin{equation}
E_2\approx a_2 (t+x)-\frac{ 2 }{3}a_2^2 (t+x)^3.
\label{txE}
\end{equation}
The electric field reaches the maximum $E_{2,m}=\sqrt{2}/3$ at $(t+x)_m=1/\sqrt{2} a$   
with the maximum width equal to $\delta (t+x)_m=1/\sqrt{2}a$.
As it is seen in Fig. \ref{FIG1} c), where we show the   local structure of the electric 
field in the wave and the corresponding time dependence of the longitudinal velocity 
of the layer emitting the wave,  spikes  of the electric field are formed 
in the reflected wave when the layer stops, 
i.e. at $v_x=0$.

\subsection{Frequency spectrum of the reflected EM radiation}
\begin{figure}[tbph]
\centering
\includegraphics[width=7cm,height=5cm]{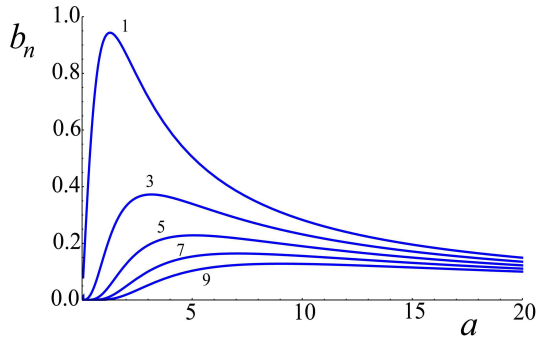} 
\caption{(Color online) Dependence of $b_n$ on the wave amplitude $a$ for $n=1,2,3,5,7,9$.}
\label{FIGII}
\end{figure}

The EM wave reflection from the electron layer accelerated by the wave  
is a simple model of a relativistic oscillating mirror. 
In the case of a linearly  polarized pulse with   $a_{0,2}(\tau) =  a_{2}\,  \sin \tau$ and $a_{0,3}(\tau)=0,$ 
the reflected periodic EM wave takes the form with  spikes shown in Fig. \ref{FIG1} b). 
It can be represented by the Fourier series  
%
\[E(\tau)=\sum_{n=1}^{\infty}b_n \sin(n \tau)\]
%
with Fourier coefficients $b_n(a)$ that vanish for  
even harmonic numbers $n$  and that can be expressed in terms of hypergeometric functions.
\[ b_n=\epsilon _{\alpha }\frac{\pi a_{2}}{1+a_{2}^{2}}\times \]
\[\left[ 
_{3}\tilde{F}_{2}\left( \left\{ \frac{1}{2},1,1\right\} ,
\left\{ \frac{3}{2}-\frac{n }{2},\frac{1}{2}+\frac{n }{2}\right\} ,
\frac{a_{2}^{2}}{1+a_{2}^{2}}\right) \right.\]
\begin{equation}
-\left._{3}\tilde{F}_{2}\left( \left\{ \frac{1}{2},1,1\right\} ,
\left\{ \frac{1}{2}-\frac{n }{2},\frac{3}{2}+\frac{n }{2}\right\} ,
\frac{a_{2}^{2}}{1+a_{2}^{2}}\right) \right].
\label{b_n}
\end{equation}
Here ${_p}\tilde{F}_{q}(\{a_p\},\{b_q\},z)$ is the regularized hypergeometric 
function equal to ${_p}{F}_{q}(\{a\},\{b\},z)/(\Gamma(b_1) ... \Gamma(b_q))/$.
 In Fig. \ref{FIGII} we plot the dependence  of   $b_n$ on the wave amplitude $a$ for $n=1,2,3,5,7,9$.

In the frame of reference where the layer 
is  on average at rest the electric field spikes  of the back reflected  
wave shown in Fig. \ref{FIG1} b) are formed at the moment when the mirror reaches 
its  maximum velocity in the backward direction. 
	The  velocity of this frame is is equal to 
\begin{equation}
v_f=\frac{1}{2\pi}\int_{-\pi}^{+\pi} \frac{a_{0,2}^2(\tau) 
+a_{0,3}^2(\tau)}{2+ a_{0,2}^2(\tau) +a_{0,3}^2(\tau)}d\tau.
\label{v-frame}
\end{equation}
%
In the case of the linearly polarized wave with $a_{0,2}(\tau)=a_{0} \sin \tau$, 
the layer moves on average with the velocity $v_f=1-1/(1+a_{2}^2/2)^{1/2}$.
The spike width and amplitude in this frame of reference changes according 
to the Lorentz transformation rules.

\subsection{Finite radiation friction force effect}

In general case, if $\epsilon_{\alpha}\ne 0$, the radiation losses lead to a finite acceleration of the layer.
Now we assume that the laser radiation has the form of a Gaussian electromagnetic  pulse with  vector potential
\[{\bf a}_{0,\perp}(x,t)={\rm exp}
\left[
-\frac{(t-x)^2}{t_{EM}^2}
\right]\]
\begin{equation}
\times \left[a_2\cos{(t-x)}{\bf e}_2+a_3\sin{(t-x)}{\bf e}_3\right].
\label{pulse1}
\end{equation}

Numerical integration of Eq. (\ref{h_equation-tau}) using relationships 
(\ref{p1_equation-tau} -- \ref{p3_equation-tau})  yields the dependence of 
the longitudinal momentum $p_1$ on the variable $\tau_{\alpha}$ 
for different values of the parameters of the electromagnetic 
pulse and of the charged layer. 

\begin{figure}[tbph]
\centering
\includegraphics[width=7cm,height=5cm]{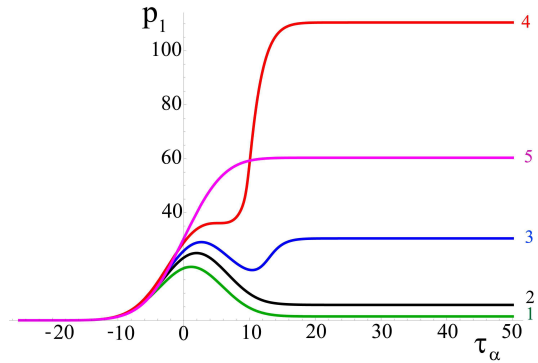} 
\caption{(Color online) Dependence of 
the longitudinal momentum $p_1$ on the variable $\tau_{\alpha}$ 
for different values of the parameter $\epsilon_{\alpha}$: 
1. $\epsilon_{\alpha}=0.03$; 2. $\epsilon_{\alpha}=0.04$; 3. $\epsilon_{\alpha}=0.045$; 
4. $\epsilon_{\alpha}=0.05$; 5. $\epsilon_{\alpha}=2.5$.}
\label{FIG2}
\end{figure}

In Fig. \ref{FIG2} we plot the longitudinal momentum $p_1$ versus $\tau_{\alpha}$ for a
circularly polarized electromagnetic pulse with amplitude equal to $a_2=a_3=a_0= 5$ and length  $t_{EM}=3 \pi$. 
The parameter $\epsilon_{\alpha}$ varies from 0.03 to 2.5.

\begin{figure}[tbph]
\centering
\includegraphics[width=7cm,height=5cm]{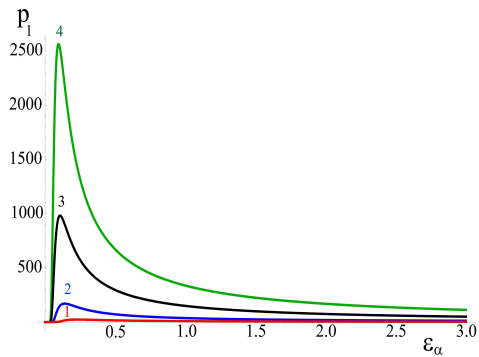} 
\caption{(Color online) Dependence of 
the longitudinal momentum $p_1$ on the parameter $\epsilon_{\alpha}$ 
for different values of the electromagnetic pulse amplitude: 
1. $a_0=1.25$; 2. $a_0=2.5$; 3. $a_0=5$; 4. $a_0=7.5$.}
\label{FIG3}
\end{figure}

As we see, in the limit of very low $\epsilon_{\alpha}$ (curves 1 and 2) the layer momentum dependence 
on $\tau_{\alpha}$ follows  
approximately according to
 Eqs. (\ref{LW1}). For larger values of $\epsilon_{\alpha}$ (curves 3 and 4) as a result of the layer 
 interaction with a finite width electromagnetic pulse the 
 momentum does not vanish at $\tau_{\alpha} \to +\infty$, i. e. the Lawson -- Woodward theorem is not valid. 
 When the parameter further increases (curve 5) the maximum
  value of the longitudinal momentum becomes lower. This fact is illustrated in Fig. \ref{FIG3}, 
  where the layer momentum dependence on $\epsilon_{\alpha}$ is shown for 
  different laser pulse amplitudes. 
  
When the  interaction of the charged layer with the electromagnetic wave occurs 
in the regime beyond the  Lawson -- Woodward theorem 
the  effects  of  the finite radiation friction force modify the electric charge dynamics 
due to its acceleration by the radiation pressure
  \cite{LandauLifshitzVol2}. 
 This is seen in the curves 3,4, and 5 in Fig. \ref{FIG2} as a  "re-acceleration"of $p_1(\tau_{\alpha})$.

\begin{figure}[tbph]
\centering
\includegraphics[width=7cm,height=5cm]{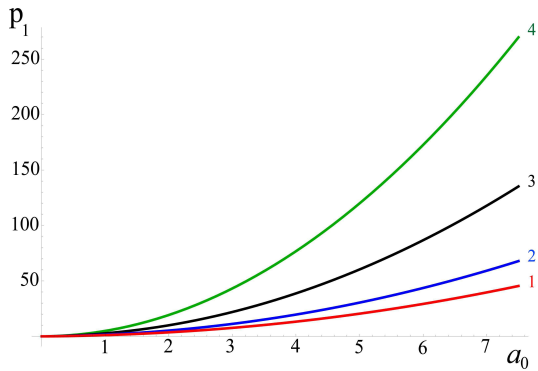} 
\caption{(Color online) Dependence of 
the longitudinal momentum $p_1$ on the electromagnetic pulse amplitude 
$a_0$  for different values of the parameter $\epsilon_{\alpha}$  : 
1. $\epsilon_{\alpha}=1.25$; 2. $\epsilon_{\alpha}=2.5$; 3. $\epsilon_{\alpha}=5$; 4. $\epsilon_{\alpha}=7.5$.}
\label{FIG4}
\end{figure}

Fig. \ref{FIG4} presents the  dependence of the layer momentum $p_1$ on the electromagnetic wave amplitude 
  for different values of the parameter $\epsilon_{\alpha}$.

\begin{figure}[tbph]
\centering
\includegraphics[width=7cm,height=7cm]{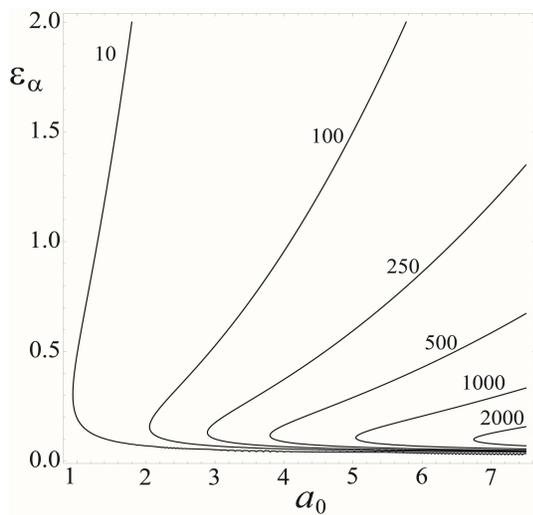} 
\caption{Isocontours of equal value of $\gamma_{\alpha}$ in the plane $\epsilon_{\alpha}, a_0$.}
\label{FIG5}
\end{figure}

The plot in Fig. \ref{FIG5} shows isocontours of equal 
value of $\gamma_{\alpha}$ in the plane $\epsilon_{\alpha}, a_0$
As we see the maximum acceleration efficiency corresponds 
to the wave amplitude of the order of $1/\epsilon_{\alpha}$ .

\section{Relativistic oscillating mirror}

The Relativistic Oscillating Mirror (ROM) concept has been proposed in 
Ref. \cite{RelOscMirr} as a mechanism of high order harmonic generation 
when an overdense plasma is irradiated by a relativistically intense laser radiation. 
The generation of high frequency radiation in thus interaction regime 
was experimentally demonstrated in Refs. \cite{Dromey_ROM}.
Within the framework of the ROM concept, 
attention is paid to 
the fact that under the laser field action the critical density region from which the light is reflected
oscillates periodically back and forth forming in other words an oscillating mirror. 
Due to the Doppler effect when the wave reflects from the relativistic mirror its frequency spectrum 
extends into the high frequency range and the wave breaks up into short wave packets. 
The reflected wave frequency is upshifted to a  range determined by a factor approximately equal to $4 \gamma_M^2$, 
where $\gamma_M$  is the relativistic gamma factor associated with the mirror motion. 
A detailed discussion of the main features  of the ROM theory 
and its experimental demonstration can be found in the review 
articles \cite{photon-rev1}. 
 A thin foil made  of two layers of electrons and ions  irradiated by a high intensity electromagnetic wave
 provides a good theoretical model elucidating the  basic features of the ROM concept. 
In this Section we assume that the ion layer is at the rest at $x_i=0$. 
When the electron layers moves with respect to the ion layer an   
electric field due to charge separation is generated equal to 
\begin{equation}
{\bf E}(x)=-\epsilon_e {\rm sgn}(x){\bf e}_1.
\label{E1}
\end{equation}

We consider an electromagnetic pulse whose form is given by 
Eq. (\ref{pulse1}), normally incident on the foil. 
 The the amplitude and the duration of this   linearly polarized short 
 pulse are  $a_0=25$ and $t_{EM}=5 \pi$, respectively. 
The ion layer is assumed to be at the rest at $x=0$. 
In the numerical integration,  in the expression for the restoring 
electric field $E(x)$,  we replace the discontinuous function ${\rm sgn}(x)$
 by  ${\rm Tanh}(x/l)$ with the plasma layer thickness equal to $l=0.01$. 
Before the laser pulse hits the target $t\to -\infty$ the electrons 
are located at $x=0$ with ${\bf p}_0=0$.

\subsection{Opaque mirror}

\begin{figure}[tbph]
\centering
\includegraphics[width=7cm,height=4.8cm]{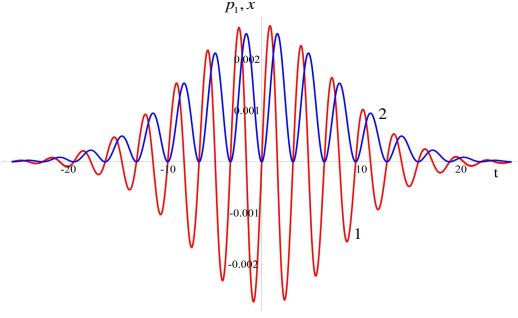} 
\caption{(Color online) Time dependence of the longitudinal electron momentum, $p_1(t)$, 
(red curve) and of  the layer coordinate, $x_1(t)$, (blue curve) for 
$a_0=25$, $t_{EM}= 5 \pi$ and $\epsilon_e=50$.}
\label{FIG6}
\end{figure}

\begin{figure}[tbph]
\centering
\includegraphics[width=7cm,height=6cm]{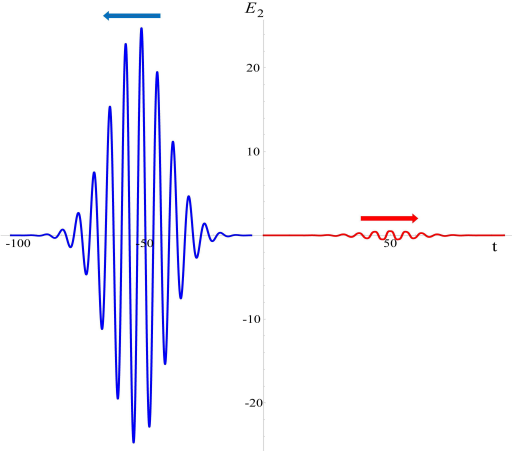} 
\caption{(Color online) Reflected, $E_2(t+x)$, (blue curve) and transmitted, $E_2(t-x)$, (red curve) waves for 
$a_0=25$, $t_{EM}= 5 \pi$ and $\epsilon_e=50$.}
\label{FIG7}
\end{figure}

 We take the dimensionless parameter $\epsilon_e$,  that  characterizes  both the radiation 
losses and the electric charge separation electric field,  equal to $\epsilon_e=50$. 
This choice corresponds to the limit when $a_0 \ll \epsilon_e$, and thus 
in this case the foil  is almost opaque for the  incident EM radiation. 
The electric charge separation field is relatively strong which results in the electron layer 
oscillations remaining in  close proximity of the ion layer. Figs. \ref{FIG6} 
and \ref{FIG7} illustrate the main features 
of the linearly polarized EM pulse interaction with the opaque foil target. 
As we see in Fig. \ref{FIG6} the electron layer  
oscillates at the front of the ion layer due to the combined effect 
of the reflected electromagnetic pulse and of the restoring 
force due to the ion layer:   the net  displacement  at the end of the pulse interaction  is 
much smaller  than the oscillation amplitude.
The average longitudinal momentum of the electron layer is  also almost zero. 
The reflected and transmitted waves plotted in Fig. \ref{FIG7} resemble 
the incident EM pulse (\ref{pulse1}). The EM wave is almost completely reflected 
with the maximum amplitude of the reflected wave equal to 24.4.
The transmitted wave calculated as the superposition of the incident wave and of 
the wave emitted forwards by the electron layer, 
which almost cancel each other,
 has its maximum amplitude equal to 0.6.

\subsection{Transparent mirror}

\begin{figure}[tbph]
\centering
\includegraphics[width=7cm,height=6cm]{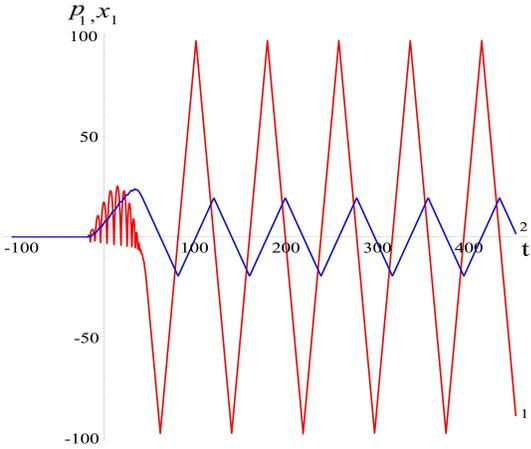} 
\caption{(Color online) Time dependence of the longitudinal electron momentum, $p_1(t)$, 
(red curve, 1) and of the layer coordinate, $x_1(t)$, (blue curve, 2) for 
$a_0=25$, $t_{EM}= 5 \pi$ and $\epsilon_e=5$.}
\label{FIG8}
\end{figure}

\begin{figure}[tbph]
\centering
\includegraphics[width=7cm,height=6cm]{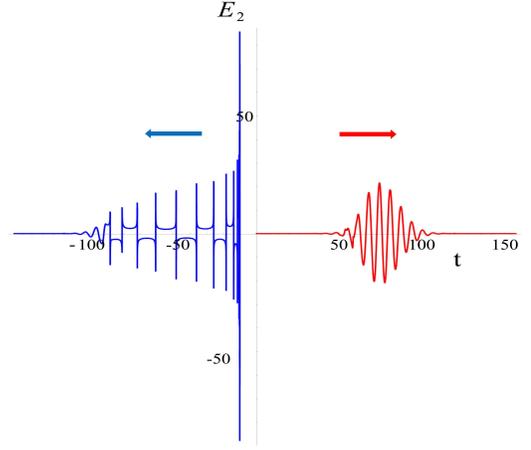} 
\caption{(Color online) Reflected, $E_2(t+x)$, 
(blue curve) and transmitted, $E_2(t-x)$, (red curve) waves for 
$a_0=25$, $t_{EM}= 5 \pi$ and $\epsilon_e=5$.}
\label{FIG9}
\end{figure}

The case of a transparent foil target with $a_0 \gg \epsilon_e$ 
is shown in Figs. \ref{FIG8} and \ref{FIG9}  for $a_0=25$ and $\epsilon_e=5$

In this regime of the EM wave interaction with the double layer target 
the radiation pressure pushes the electron layer forwards. 
The interaction is nonadiabatic with respect 
to the longitudinal ``sawtooth'' oscillation excitation which are seen in the longitudinal electron 
momentum and coordinate dependence on time presented in Fig. \ref{FIG8}. 
Similar oscillations have been noticed in Ref. \cite{MikhPRL}.
In contrast to the opaque case, the net layer displacement at the end 
of the interaction of the pulse with the layer 
is not small and provides the initial condition for the  ``sawtooth'' oscillations 
that  are a periodic sequence of hyperbolic 
motions  of the electric charge in  the homogeneous electric field \cite{LandauLifshitzVol2} 
due to the ion layer. 
Within an oscillation half cycle the electron layer 
momentum depends   on time as
\begin{equation}
p_1(t)=p_m-\epsilon_e t
\label{pmt}
\end{equation}
 in the time interval $0<t<t_m$, where $t_m$ is the half-cycle duration equal to $2 p_m/\epsilon_e$. 
 The time dependence of the layer coordinate is given by 
\begin{equation}
x_1(t)=\frac{1}{\epsilon}\left[\sqrt{1+p_m^2}-\sqrt{1+(p_m-\epsilon_e t)^2}\right]
\label{xmt}
\end{equation}
The maximum of the electron layer momentum $p_m$ and the maximum 
of the layer displacement $x_m$ are related to each other as
\begin{equation}
p_{1,m}=\sqrt{(1+\epsilon_e x_{1,m})^2-1}.
\label{xmpm}
\end{equation}
In order to find $x_{1,m}$ in the limit $a_0 \gg \epsilon_e$ we can use expression (\ref{displace}), 
which for the Gaussian linearly polarized EM pulse (\ref{pulse1}) yields 
\begin{equation}
x_{1,m}=a_0^2
\frac{\left[1-\exp\left(-t_{EM}^2/2 \right)\right]\sqrt{\pi} t_{EM}}
{2^{5/2}}.
\label{xmtEM}
\end{equation}
The condition of nonadiabatic interaction is $t_m>t_{EM}$.
The electron kinetic energy found from Eqs. (\ref{xmpm}) and (\ref{xmtEM}) is given by 
\begin{equation}
\frac{{\cal E}_e}{m_ec^2}=
\gamma_e-1=\epsilon_e x_{1,m}\approx \sqrt{\frac{\pi}{32}}a_0^2 \epsilon_e t_{EM},
\label{xmtEM}
\end{equation}
which for $a_0^2 \epsilon_e t_{EM}\gg 1$ is well above the quiver energy of an electron moving in the EM wave.
 The excitation of the sawtooth oscillations can be regarded as the efficient collisionless heating of  the electrons. 
This in fact can be an underlying mechanism  of the electron energization during high intensity laser radiation 
interaction with a thin foil target observed in 
the computer simulations presented in Ref. \cite{IONS-UL} (see Fig. 1 (b) therein).

The transmitted and reflected waves shown in Fig. \ref{FIG9} have  approximately of the same amplitude level 
 because the receding relativistic mirror  becomes less transparent while it is accelerated in  the forward 
 direction \cite{photon-rev3, MacchiPRLReflectivity, BulanovOptimalShape}.
This  results in a relative  enhancement of the reflected wave amplitude. 
The spectra of the reflected and transmitted radiation contain high order harmonics. 
The reflected wave has the  form of ultrashort spikes. 
The distance between them corresponds to the stretched  wavelength of the incident light due to 
the double Doppler effect, because part of the wave interaction with the oscillating electron layer 
occurs under the conditions of  reflection from a receding mirror. 
We note that the strongest  spike at the rear of the reflected 
pulse is formed due to interaction with the sawtooth oscillations.


\section{Relativistic flying mirror}

A method to generate high frequency radiation based on the concept of the 
{\it Relativistic Flying Mirror} (RFM) considers a
thin plasma shell travelling close to the speed of light as
a relativistic mirror. The reflected light undergoes frequency upshift, compression and intensification
due to a relativistic double Doppler effect. Various schemes
were described 
\cite{photon-rev3, BulanovRMP, Bulanov2006, Kulagin2007, Kulagin2013, Kagami, BrII2012, Lobet} 
and experimentally demonstrated
\cite{Kando2007, Foil3} as a proof of the feasibility of this concept.

\subsection{The shape of a pulse reflected from a relativistic flying mirror }
Using a double layer thin foil target as a RFM model,  
we consider the configuration of two counter propagating pulses. 
The first EM pulse driver pushes the 
electron layer forwards with relativistic velocity. The second pulse is relatively week 
and propagates in the opposite direction. 
As a result of its head-on collision with the RFM  a portion of the photons from this pulse is back reflected. 
This process is accompanied by the frequency upshifting of
the reflected photons and by the modulation of the reflected pulse.
When the ponderomotive force of the driver EM pulse is substantially 
larger than the force from the electric field due to 
the electric charge separation, i. e. when 
$\epsilon_e \ll a_0$, the motion of the relativistic electron layer 
can be described by Eqs. (\ref{eq:h} -- \ref{LW4}). 
In the case when the electron layer is accelerated by a 
linearly polarized EM pulse, as  analysed in Ref. \cite{photon-rev3}, the phase of the reflected
 part of the weaker  EM wave is given by 
\begin{equation}
\psi_r(u)=\omega_s\left(u +\frac{a_0^2}{2}u-\frac{a_0^2}{4 \omega} \sin 2\omega u\right)
\label{reflphase}
\end{equation}
with $u=t-x$ and $\omega_s$   the frequency of the EM source pulse. 
The reflected pulse frequency given by a derivative of the phase $\psi_r$ with respect to time is 
\begin{equation}
\omega_r(u)=\omega_s\left(1 +a_0^2 \sin^2\omega u\right)
\label{reflphase}
\end{equation}
The frequency upshifting factor $g=\omega_r/\omega_s$ depends 
on the longitudinal velocity of the mirror, $v_1=p_1/\gamma$ as
\begin{equation}
g=\frac{\gamma+p_1}{\gamma-p_1}.
\label{gfactor}
\end{equation}
If the source pulse frequency is equal to the driver pulse frequency, $\omega_s=\omega=\omega_0$, 
the reflected pulse frequency, $\omega_r=\omega_0(1 +a_0^2\sin^2\omega_0 u)$, changes 
from $\omega_0$ to $\omega_0 (1+a_0^2)$. 
The wave amplitude is modulated accordingly. 
The reflected radiation consists  of a sequence of short high frequency pulses. 

\begin{figure*}[tbph]
\centering
\includegraphics[width=12cm,height=9cm]{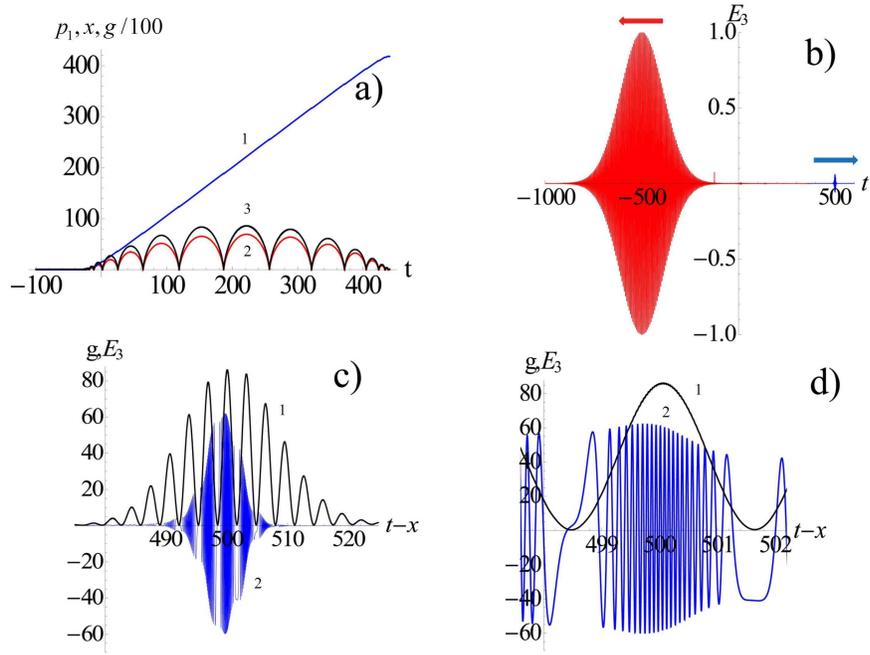} 
\caption{(Color online) a) Time dependence of the longitudinal electron layer momentum, $p_1(t)$, (red curve),  
of the layer coordinate, $x_1(t)$,
(blue curve) and of the factor $g(t)$ (black curve) for the driver EM pulse with 
$a_2 = 15$, $t_{EM,d} = 5 \pi$ and $\epsilon_e= 0.1$.
Counterpropagating source pulse:
b) Reflected, $E_3(t-x)$, (blue curve) and transmitted, $E_3(t + x)$, (red curve) waves for 
$a_3=1$, $t_{EM,s}= 150 \pi$ and $\epsilon_e=0.1$.
c) Reflected pulse (blue curve) and frequency upshifting factor $g(t)$ (black curve).
d) Close up of the reflected pulse (blue curve) and frequency upshifting factor $g(t)$ (black curve).}
\label{FIG10}
\end{figure*}

Fig. \ref{FIG10} shows the results  of the numerical integration of Eqs. (\ref{p1_equation} -- \ref{p3_equation}). 
The Gaussian EM pulse driver is linearly polarized with $a_2=15, \qquad a_3=0$ and $t_{EM,d}=5 \pi$. 
The source EM pulse is linearly polarized in the perpendicular plane, 
with $a_2=0, \qquad a_3=1$ and $t_{EM,s}=150 \pi$. 

In Fig. \ref{FIG10} a) we plot  the time dependence of the longitudinal momentum, 
$p_1(t)$,  of the electron layer  (red curve), 
 of  the layer coordinate, $x_1(t)$,
(blue curve) and of the factor $g(t)$ (black  curve) for the driver EM pulse 
with $a_2 = 15$, $t_{EM,d} = 5 \pi$ and $\epsilon_e= 0.1$.
The electron layer,  while oscillating back and forth, moves on average  forwards with a  relativistic velocity. 
The frequency upshifting factor $g$ oscillates synchronously 
with the layer momentum $p_1$. According to expressions (\ref{LW1} -- \ref{LW4}) and (\ref{gfactor}) the 
factor $g$ and the layer longitudinal momentum are related  in the limit $a_0 \gg \epsilon_e$ as 
$g=1+2 p_1$, i.e. the factor $g$ scales with the layer energy as $g= \sim 2\gamma_e$. 
For the chosen EM pulse driver  amplitude equal to $15$ the maximum value of the factor $g$ is 226.

\begin{figure}[tbph]
\centering
\includegraphics[width=6cm,height=8cm]{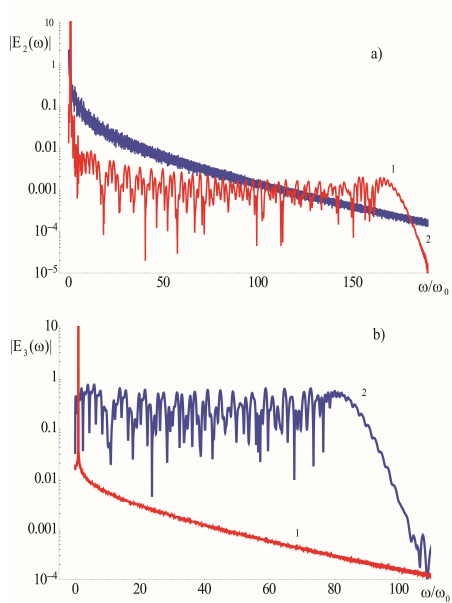} 
\caption{(Color online) The frequency spectrum of the driver and source pulses.
a) The dependence of the absolute value of the Fourier transform of the 
$E_2$ component of the electric field, corresponding to the incident and transmitted electromagnetic 
of the driver pulse.
b) The dependence 
of the absolute value of the Fourier transform of the 
$E_3$ component of the electric field, which corresponds to the incident and reflected electromagnetic 
of the source pulse. }
\label{HHG}
\end{figure}

In Fig. \ref{HHG} we present the frequency spectrum of the driver and source pulses. 
Fig. \ref{HHG} a) shows the dependence of the absolute value of the Fourier transform of the 
$E_2$ component of the electric field, corresponding to the incident and transmitted electromagnetic 
of the driver pulse. In Fig. \ref{HHG} b) we plot the dependence 
of the absolute value of the Fourier transform of the 
$E_3$ component of the electric field, which corresponds to the incident and reflected electromagnetic 
of the source pulse. The spectrum of the reflected radiation is enriched by the high order harmonics.
It has a form of the plateau, which extends to the value of the order of $\omega_{\max}\approx \omega_0 g$

\begin{figure}[tbph]
\centering
\includegraphics[width=7cm,height=8cm]{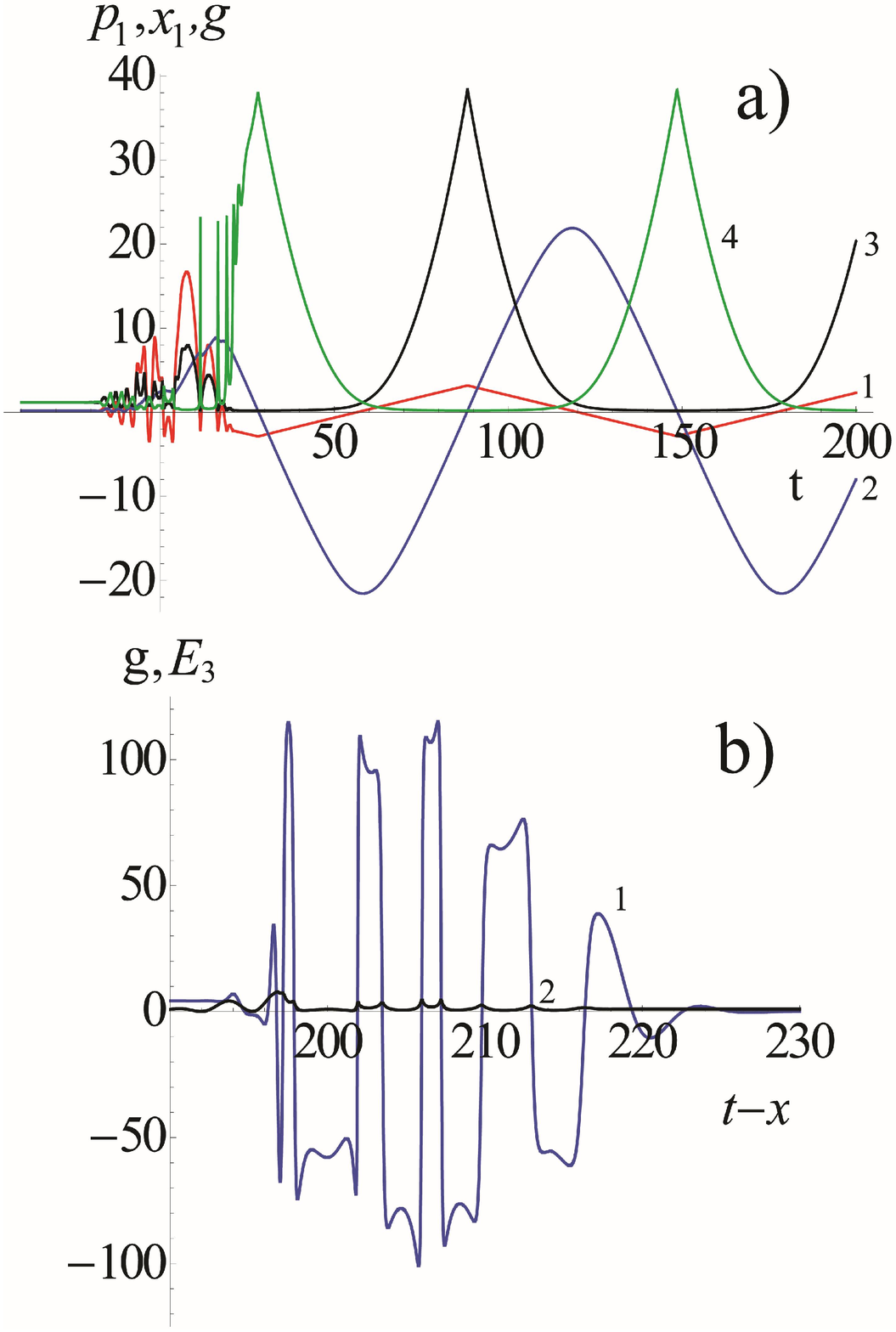} 
\caption{(Color online) Counterpropagating driver and source pulses of equal amplitude and duration. 
a) Time dependence of the longitudinal electron layer momentum, $p_1(t)$, (red curve), 
the layer coordinate, $x_1(t)$,
(blue curve), the factors $g_{+}(t)$ (black curve) and $g_{-}(t)$ (green curve), 
for the driver EM pulse with $a_2 = 15$, $a_3=15$, $t_{EM,d} =t_{EM,s}= 3 \pi$ and $\epsilon_e= 0.1$.
b) Reflected pulse $E_3(t-x)$(blue curve) and frequency upshifting factor $g_{+}(t)$ (black curve).}
\label{FIG11}
\end{figure}

The reflected EM pulse as seen in Fig. \ref{FIG10} b) is approximately 
shorter   by a factor $g=226$  than the source pulse  incident on the foil. 
Fig. \ref{FIG10} c) shows that the reflected wave breaks up into a train of high frequency pulses, 
which are frequency modulated (see Fig. \ref{FIG10} d)),
i.e. in general the frequency upshifting and shortening  of the  reflected pulse 
is accompanied by the generation of  high order harmonics.

The amplitude of the reflected EM pulse is proportional to the amplitude of the incident radiation, $a_s$, 
times the factor $g$ and times 
the reflection coefficient $\rho$. The reflection coefficient can be found as in Refs. 
\cite{Foil1, photon-rev3, MacchiPRLReflectivity, BulanovOptimalShape, BrII2012}. 
In the frame of reference co-moving with the electron 
layer where the longitudinal momentum component vanishes, $p_{1}=0$, the equation for the electric field, 
${\bf E}={\bf E}_{0}+{\bf E}_{e,l}$, according to Eqs. (\ref{E_equation}) and (\ref{E_ext}) 
can be written in the form 
\begin{equation}
\label{reflect}
{\bf E}'={\bf E}'_{0}+\frac{2\pi e n l}{c}{\bf v}'_{\perp},
\end{equation}
where a prime denotes the electric field and the electron velocity in the co-moving frame of reference 
and ${\bf v}'_{\perp}=v'_2 {\bf e}_2+v'_3 {\bf e}_3$.
Here we use dimensional variables, i.e. $2\pi e n l$ instead $\epsilon_e$, 
in order to clearly show that the areal charge density, $enl$, is  Lorentz invariant while 
the electric field and the electron velocity are not invariant. In the 
head-on collision configuration of the EM pulse
interaction with the electron layer when $v_{1}<0$ the electric field in the boosted frame 
is larger than that in the laboratory frame of reference by a factor of 
 $\sqrt{(1+|v_{1}|)/(1-|v_{1}|)}\approx 2 \gamma$. 

Since in any frame of reference the electron velocity cannot exceed the speed of light in vacuum, 
there are two limiting cases depending on the value of 
$E'_0/2\pi e n l\approx 4 \gamma^2a_s/\epsilon_e$. In the case of a weak EM wave, 
when this ratio is much smaller  than unity, 
from Eq. (\ref{reflect}) it follows that the 
amplitudes of the incident, $E'$, and reflected, $2\pi e n l{\bf v}'_{\perp}/c$, 
waves are almost equal to each other, 
i.e. the reflection coefficient is
of the order of unity. In the opposite limit, when $E'_0/2\pi e n l\gg 1$, 
the amplitude of the reflected EM wave 
in the boosted frame of reference 
is of the order of $2\pi e n l$. This yields a constraint on the upper 
limit of the EM radiation intensity measured 
in the laboratory frame of reference, 
when the wave is reflected by a thin electron layer of areal density $nl$ 
moving with relativistic gamma-factor $\gamma$, 
as in the above considered case or 
in the flying mirror configuration discussed in Ref. \cite{Kulagin2007}: $I_r \le 16 \pi c (e n l)^2 \gamma^2$.  
For example,
for an 10$^{-2}\mu$m, $n=10^{23} {\rm cm}^{-3}$ electron layer moving with the gamma-factor equal to 10$^2$ 
this yields $I_r \approx 5\times 10^{25}$W/cm$^2$.

In Fig. \ref{FIG11} we illustrate the regime when two EM pulses with  equal 
amplitudes  and perpendicular polarizations
  interact with a thin foil target. 
The amplitudes of the driver and source 
pulses are equal to $a_2=15$, $a_3=15$, $t_{EM,s}=t_{EM,d}=3 \pi$ and $\epsilon_e=0.1$. 
In Fig. \ref{FIG11} a), we plot the  time dependence of the electron layer coordinate $x(t)$, 
momentum component $p_1(t)$ and of the frequency upshifting factors $g_{+}$ and $g_{-}$, 
for the waves reflected to the right and to the left hand side directions, respectively. 
As seen, during the interaction of the two EM pulses colliding head-on 
the electron layer undergoes irregular jigglings. 
The frequency upshifting factors are not as large 
as in the previous 
case presented in Fig. \ref{FIG10}. The reflected EM wave shown in Fig.\ref{FIG11} 
b) has  a non-sinusoidal form  and is much less regular than in the case of  a  
source pulse with a finite but not too 
large amplitude described by  Figs. \ref{FIG11} c) and d).

\subsection{Head on interaction of an EM pulse with an electron layer in the regime of sawtooth oscillations}
\begin{figure}[tbph]
\centering
\includegraphics[width=6cm,height=7cm]{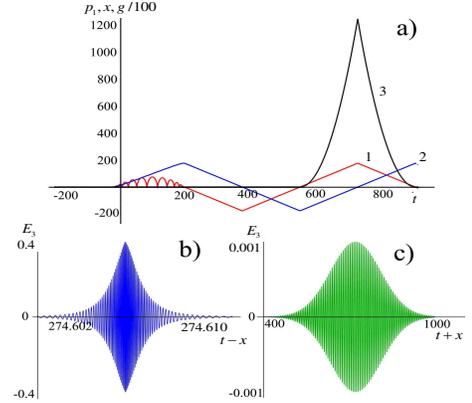} 
\caption{(Color online) 
a)Time dependence of the longitudinal electron layer momentum, $p_1(t)$, 
(red curve), of the layer coordinate, $x_1(t)$,
(blue curve) and  of the  factor $g(t)$ divided by 100 (black curve) 
for a driver EM pulse with $a_2 = 25$, $t_{EM,d} = 5 \pi$ and $\epsilon_e= 1$.
b) Reflected pulse $E_3(t-x)$(blue curve).
c) Incident source pulse $E_3(t+x)$(green curve)
}
\label{FIGXI}
\end{figure}

As  noticed above (\ref{gfactor}), if the electron layer is driven by an 
electromagnetic wave with amplitude $a_0$ 
the frequency upshifting factor for a counterpropagating pulse cannot exceed the  value $g \le (1+a_0^2)$. 
However, the $g$ factor can be substantially enhanced by  imposing a  delay between 
the driver and the source pulses in a such a way 
that the counterpropagating source pulse gets reflected by  the electron layer  
in  the phase when the layer undergoes  
``sawtooth'' oscillations. 
 According to Eq. (\ref{xmtEM}) the mirror Lorentz factor scales as $a_0^2 \epsilon_e t_{EM}$, 
 i. e. the the frequency upshifting factor 
 may be of the order of $\approx 4 a_0^4 \epsilon_e^2 t_{EM}^2$. 
In other words the longitudinal velocity of the electron layer during 
the phase of ``sawtooth'' oscillations, i.e. after the 
 end of the driver EM pulse,
is substantially larger  than the velocity of the oscillations driven 
by the ponderomotive force as clearly seen in Fig. \ref{FIG8}. 
By choosing the delay time between the driver 
and the counter propagating source pulse in a such way that the source pulse 
collides with the electron layer at the ``sawtooth'' 
oscillation phase, we can provide conditions for a 
much higher  frequency upshifting and intensification of the back-reflected 
radiation in the regime as shown  in Fig. \ref{FIGXI}.  
The source pulse is polarized in the plane perpendicular to the driver pulse polarization plane. 
Its amplitude is equal to 0.001 and its width is $t_{EM,s}=40 \pi$. 
The delay time between the driver and source pulses is  $230 \pi$. In Fig. \ref{FIGXI} a) 
the time dependence of the longitudinal electron layer momentum, $p_1(t)$, (red curve), 
the layer coordinate, $x_1(t)$,
(blue curve) and the divided by 100 factor $g(t)$ (black curve) for the driver EM pulse with $a_2 = 25$, 
$t_{EM,d} = 5 \pi$ and $\epsilon_e= 1$ are shown. 
The frequency upshifting factor reaches its  maximum $\approx 1.2 \times 10^5$ at $t=725$. 
The reflected source pulse shown in Fig. \ref{FIGXI} b) has amplitude $E_3=0.4$ approximately 
$4\times 10^3$ times larger  than  that of the incident wave (see Fig. \ref{FIGXI} c)). 
Its width and wavelength are shortened by a factor $g$. The form  of the reflected EM pulse resembles that
of the frequency upshifting factor $g(t)$ (see Figs. \ref{FIGXI}) a) and b)).
 
\section{Ion acceleration}

In Ref. \cite{IONS} the radiation pressure exerted by an ultraintense electromagnetic pulse on a quasineutral plasma 
foil has been proposed as a very efficient acceleration mechanism capable of providing ultrarelativistic ion beams. 
In this radiation pressure dominant acceleration (RPDA) regime, the ions move forward 
under the push of the pulse pressure 
with almost the same velocity as the electrons. A fundamental feature of this 
acceleration process is its high efficiency, 
as the ion energy per nucleon turns out to be proportional in the ultrarelativistic 
limit to the electromagnetic pulse energy.

Recently the RPDA regime of laser ion acceleration has attracted great attention e.g., 
see review articles \cite{ion-rev}. In Ref. \cite{RT-PB} the stability of the accelerated foil has been analyzed. 
A foil accelerated to relativistic energies by a laser pulse can also act as a relativistic flying mirror 
for frequency upshift and intensification of a reflected counterpropagating light beam \cite{kagami}. 
An indication of the effect of the radiation pressure on bulk target ions is obtained in experimental 
studies of thin solid targets irradiated by ultraintense laser pulses \cite{SKAR}.
          
 Below we consider a double layer (ion and electron) thin foil target irradiated by the EM radiation. 
For the sake of simplicity we assume that the electron layer motion is described 
by Eqs. (\ref{p1_equation} -- \ref{p3_equation}).
In the ion layer equations of motion we neglect its interaction with the EM wave 
retaining only the electrostatic force due to the electric 
field produced by the electron layer. The electrostatic approximation for the ion layer motion can be used provided 
the parameter $eE/m_i \omega c$ is small, i.e., in the case 
of a   one micron wavelength laser,  for a  light intensity  below $\approx 10^{24}$W/cm$^2$. 
At this limit the classical electrodynamics paradigm 
must be changed  and  quantum effects  must be included \cite{QEDint}.

\begin{figure*}[tbph]
\centering
\includegraphics[width=12cm,height=9cm]{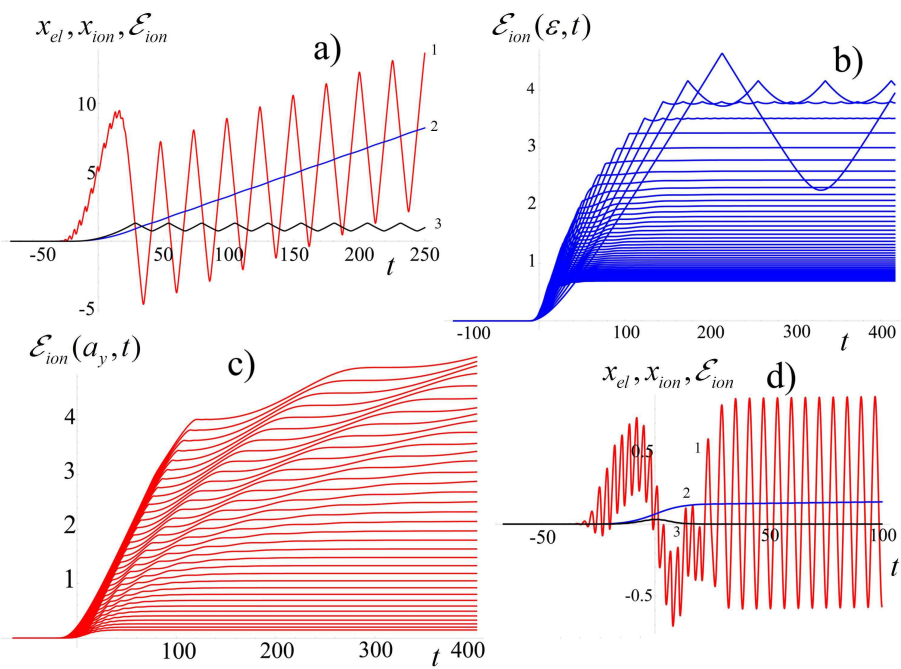} 
\caption{(Color online) Ion acceleration by the radiation pressure. 
a) Time dependence of the the electron (red curve) and ion (blue curve) 
layer coordinates and of the ion energy $m_{i}c^2 (\gamma_i-1)$ (black) 
for an EM pulse with $a_2 = 10$, $t_{EM} =5 \pi$ and $\epsilon_e= 1.5$.
b) Normalized  ion energy $\gamma_i-1$ v.s. time for $a_2=400$ and the parameter $\epsilon_e$ varying from 45 to 250 
from bottom to top with the step equal to 5.
c) Normalized  ion energy $\gamma_i-1$ v.s. time for $\epsilon_e=45$ and the EM pulse amplitude $a_2$ varying 
from bottom to top  from 100 to 450 with the step equal to 10.
d) Time dependence of the the electron (red curve) and ion (blue curve) layer coordinates 
and of the ion energy $m_{i}c^2 (\gamma_i-1)$ (black) for the case without radiation friction.}
\label{FIGXIV}
\end{figure*}

The results of the numerical integration of  the equations of motion of the electron and ion layers irradiated 
by a strong EM wave are shown in Fig. \ref{FIGXIV}.
Figure  \ref{FIGXIV} a) presents typical regimes of   ion acceleration for 
a linearly polarized electromagnetic pulse with  amplitude $a=10$ and duration $t_{EM}=5 \pi$ 
interacting with  a foil with  $\epsilon_e=1.5$. At the initial stage $-25<t<25$ the time  dependence of the 
ion and electron coordinates corresponds to a strong charge separation.
 As   seen in Fig. \ref{FIGXIV} a), the electron layer pushed by the radiation pressure 
 of the EM wave pulls the ion layer. 
Then both layers move forwards with  the same average velocity  and with the electron 
layer moving back forth around the ion layer performing 
 sawtooth oscillations. This phenomenon can explain the efficient electron heating during 
 the RPD ion acceleration observed in the PIC simulations 
presented in Ref. \cite{IONS-UL}. 
In addition these oscillations cause   oscillations  of the  ion energy $\gamma_i-1$ around its average value. 

The parametric dependence on the EM pulse amplitude  and on  the target surface density 
of the  energy of the accelerated ions is 
 illustrated in Figs. \ref{FIGXIV} b) and c).  The dependence on time of the ion energy for different values 
 of the EM pulse amplitude and varying parameter $\epsilon_e$ is not monotonic as 
seen in Figs. \ref{FIGXIV} b) and c) and  can be explained by the sawtooth oscillations 
of the electron and ion layers. 
A  finite value of the parameter $\epsilon_e$ provides an 
efficient coupling between the EM pulse and the electron-ion foil target. In the case 
without radiation friction shown in Fig.  \ref{FIGXIV} d) 
the ion acceleration is less efficient.

\section{Conclusions, Discussions, Main Results}

The  theory of the interaction of relativistically strong electromagnetic fields   
with foil targets used in the present paper 
is based on the thin layer model of the one-dimensional electrodynamics of charged particles. 
 It  describes the  1D   motion  of electric charges in the self-consistent electromagnetic field incorporating 
 the  charge self-action or, in other words, the  effect of the radiation friction force.

Within this framework,  the generation of  high order harmonics in the relativistic regime 
occurs through  the electromagnetic wave 
reflection, or  collective backward scattering also called nonlinear collective 
Thomson scattering,  at the electron layer  driven  
by the electromagnetic wave.  The back reflected radiation takes the  form of a train 
of ultrashort single-cycle electromagnetic pulses 
that are formed at the moment of maximal negative velocity of the layer.

The radiation scattered by the thin foil target in the backward and that scattered 
in the forward direction have different frequency spectra.

In the nonadiabatic regime of interaction with the ion-electron layer 
target a short electromagnetic pulse excites 
relatively low frequency sawtooth oscillations with   amplitude 
substantially larger  than the amplitude of the  oscillations of 
the electron layer driven by the electromagnetic pulse.
These  sawtooth oscillations provide a  mechanism of collective electron heating. 
They also generate extremely short spikes in the reflected EM wave.

In the configuration when two electromagnetic beams irradiate the thin foil target the electron 
layer driven by strong enough electromagnetic wave plays  the  role 
of a relativistic flying mirror for the second pulse. The  electromagnetic 
radiation reflected from the relativistic mirror counterpropagating
is intensified and its frequency is substantially increased. 
The reflected radiation takes the  form of a sequence of the high frequency 
short bunches of electromagnetic radiation.

Since the longitudinal velocity of the electron layer during the phase 
of the ``sawtooth'' oscillations is substantially 
larger than the velocity of  the oscillations driven within the electromagnetic 
pulse driver, by choosing the delay time between the driver 
and the counter propagating source pulse in a such way that the source pulse 
collides with the electron layer at the ``sawtooth'' 
oscillation phase, we can provide conditions for  high frequency upshifting and 
intensification of the back-reflected radiation. 
 
Under the radiation pressure of the electromagnetic wave the electron layer 
becomes separated from the ion layer 
that moves  in the electric field  due the charge separation. 
As a result,  while the electron layer undergoes back and forth 
sawtooth oscillations around the ion layer,  
on average both layers  move together. The ion acceleration rate grows higher with  higher  amplitude 
of the incident electromagnetic wave. It also depends on the radiation friction 
which is responsible for the  coupling of the electromagnetic field with the electron layer because 
it provides the wave back scattering and thus the momentum transfer from the 
electromagnetic field to  the charge particles. 
If the radiation friction force effects are not taken into account 
the ion acceleration rate is substantially lower for the same 
electromagnetic pulse amplitude.

\section*{Acknowledgments}
The authors would like to thank for discussions T. M. Jeong, C. M. Kim, G. Korn, 
V. V. Kulagin, T. Levato,  D. Margarone, 
N. N. Rosanov,  H. Suk, A. Zhidkov. We appreciate support from the NSF under Grant No. PHY-0935197 and the Office of Science of the US DOE under Contract No. DE-AC02-05CH11231 and No. DE-FG02-12ER41798.


\begin{thebibliography}{90}

\bibitem{ion-rev} M. Borghesi, J. Fuchs, S. V. Bulanov, A. J. MacKinnon, 
P. K. Patel, and M. Roth, Fus. Sci. Techn. {\bf 49}, 412 (2006); 
H. Daido, M. Nishiuchi, and A. S. Pirozhkov, Rep. Prog. Phys. {\bf 75}, 056401  (2012);
A. Macchi, M. Borghesi, and M. Passoni, Rev. Mod. Phys. {\bf 85}, 751 (2013). 

\bibitem{ele-rev} E. Esarey, C. B. Schroeder, and W. P. Leemans, Rev. Mod. Phys. {\bf 81}, 1229  (2009).

\bibitem{photon-rev1} U. Teubner and P. Gibbon, Rev. Mod. Phys. {\bf 81}, 445 (2009); 
F. Krausz and M. Ivanov, Rev. Mod. Phys. \textbf{81}, 163 (2009).

\bibitem{photon-rev2} S. Corde, K. Ta Phuoc, G. Lambert, R. Fitour, V. Malka, A. Rousse, A. Beck, E. Lefebvre, 
Rev. Mod. Phys. \textbf{85}, 1 (2013).

\bibitem{photon-rev3} S. V. Bulanov, T. Zh. Esirkepov, A. S. Pirozhkov, and N. N. Rosanov, 
Physics Uspekhi {\bf 56}, 429 (2013).

\bibitem{MTB}G.~Mourou, T. Tajima, and S. V. Bulanov, Rev. Mod. Phys. \textbf{78}, 309 (2006).

\bibitem{fund-rev} M. Marklund and P. Shukla, Rev. Mod. Phys. {\bf 78}, 591 (2006); 
A. Di Piazza, C. Muller, K. Z. Hatsagortsyan, and C. H. Keitel, Rev. Mod. Phys. \textbf{84}, 1177 (2012).

\bibitem{astro-rev} B. A. Remington, D. Arnett, R. P. Drake, and H. Takabe, Science {\bf 284}, 1488 (1999); 
B. Remington, R. P. Drake, and D. Ryutov, Rev. Mod. Phys. {\bf 78}, 755 (2006); 
S. V. Bulanov, T. Zh. Esirkepov, D. Habs, F. Pegoraro, and T. Tajima, Eur. Phys. J. D {\bf 55}, 483 (2009).

\bibitem{Foil1}V. A. Vshivkov, N. M. Naumova, F. Pegoraro, S. V. Bulanov, 
Phys. Plasmas {\bf 5}, 2727 (1998).

\bibitem{IONS}T. Esirkepov, M. Borghesi, S. V. Bulanov, G. Mourou,
and T. Tajima, Phys. Rev. Lett. {\bf 92}, 175003 (2004).

\bibitem{IONS-UL} S. V. Bulanov, E. Yu. Echkina, T. Zh. Esirkepov, I. N. Inovenkov, M.
Kando, F. Pegoraro, and G. Korn, Phys. Rev. Lett. {\bf 104}, 135003 (2010); 
S. V. Bulanov, E. Yu. Echkina, T. Zh. Esirkepov, I. N. Inovenkov, M. Kando,
F. Pegoraro, and G. Korn, Phys. Plasmas {\bf 17}, 063102 (2010).

\bibitem{AAMS} H. K. Avetissian, A. K. Avetissian, G. F. Mkrtchian, 
and Kh.V. Sedrakian, Phys. Rev. STAB {\bf 14}, 101301 (2011). 

\bibitem{MacchiPRLReflectivity}A.~Macchi, S. Veghini, and F. Pegoraro, 
Phys. Rev. Lett. {\bf 103}, 085003 (2009);  
S. V. Bulanov, T. Zh. Esirkepov, Y. Hayashi, M. Kando, H. Kiriyama, 
J. K. Koga, K. Kondo, H. Kotaki, A. S. Pirozhkov, 
S. S. Bulanov, A. G. Zhidkov, P. Chen, D. Neely, Y. Kato, 
N. B. Narozhny, and G. Korn, Nucl. Instr. Meth. Phys. Res. A  {\bf 653}, 153 (2011). 

\bibitem{BulanovOptimalShape}S. S.~Bulanov, C. B. Schroeder, E. Esarey,  
and W. P. Leemans, Phys. Plasmas {\bf 19}, 093112 (2012).

\bibitem{RelOscMirr}S. V.~Bulanov, N. M. Naumova, and F. Pegoraro, Phys. Plasmas {\bf 1}, 745 (1994).

\bibitem{Pirozhkov2006} A. S. Pirozhkov, S. V. Bulanov, T. Zh. Esirkepov, M. Mori, A. Sagisaka and H. Daido, 
Phys. Lett. A {\bf 349}, 256 (2006);
A. S. Pirozhkov, S. V. Bulanov, T. Zh. Esirkepov, M. Mori, A. Sagisaka and H. Daido,
Phys. Plasmas {\bf 13}, 013107 (2006).

\bibitem{MikhPRL}J. M. Mikhailova, M. V. Fedorov, N. Karpowicz, 
P. Gibbon, V. T. Platonenko, A. M. Zheltikov, and F. Krausz, 
Phys. Rev. Lett. {\bf 109}, 245005 (2012).

\bibitem{Dromey_ROM} B. Dromey, M. Zepf, A. Gopal, K. Lancaster, M. S. Wei, K. Krushelnick, M. Tatarakis, N. Vakakis, S. Moustaizis, R. Kodama, M. Tampo, C. Stoeckl, R. Clarke, H. Habara, D. Neely, S. Karsch, and P. Norreys, Nat. Phys. \textbf{2}, 456 (2006); B. Dromey, S. Kar, C. Bellei, D. C. Carroll, R. J. Clarke, J. S. Green, S. Kneip, K. Markey, S. R. Nagel, P. T. Simpson, L. Willingale, P. McKenna, D. Neely, Z. Najmudin, K. Krushelnick, P. A. Norreys, and M. Zepf, Phys. Rev. Lett. \textbf{99}, 085001 (2007); B. Dromey, D. Adams, R. Hšrlein, Y. Nomura, S. G. Rykovanov, D. C. Carroll, P. S. Foster, S. Kar, K. Markey, P. McKenna, D. Neely, M. Geissler, G. D. Tsakiris, and M. Zepf, Nat. Phys. \textbf{5}, 146 (2009); B. Dromey, S. Rykovanov, M. Yeung,	 R. Hšrlein, D. Jung, D. C. Gautier, T. Dzelzainis, D. Kiefer,	 S. Palaniyppan,	 R. Shah, J. Schreiber, H. Ruhl, J. C. Fernandez, C. L. S. Lewis,	 M. Zepf, and B. M. Hegelich, Nat. Phys. \textbf{8}, 804 (2012).

\bibitem{UofM_HHG} F. Dollar, P. Cummings, V. Chvykov, L. Willingale, M. Vargas, V. Yanovsky, C. Zulick,
A. Maksimchuk, A. G. R. Thomas, and K. Krushelnick, Phys. Rev. Lett. \textbf{110}, 175002 (2013).

\bibitem{BulanovRMP}S. V. Bulanov, T. Zh. Esirkepov, and T. Tajima, 
Phys. Rev. Lett. \textbf{91}, 085001 (2003).

\bibitem{Bulanov2006} S. S. Bulanov, 
T. Zh. Esirkepov, F. F. Kamenets, and F. Pegoraro, 
Phys. Rev. E \textbf{73}, 036408 (2006); 
S. S. Bulanov, 
A. Maximchuk, C. B. Schroeder, A. G. Zhidkov, E. Esarey, and W. P. Leemans, 
Phys. Plasmas \textbf{19}, 020702 (2012). 

\bibitem{Kulagin2007} 
V. V. Kulagin, V. A. Cherepenin, and H. Suk,
Appl. Phys. Lett. {\bf 85}, 3322 (2004);
V. V. Kulagin, 
V. A. Cherepenin, M. S. Hur, and H. Suk, 
Phys. Plasmas \textbf{14}, 113101 (2007); 
D. Habs, 
M. Hegelich, J. Schreiber, M. Gross, A. Henig, D. Kiefer, and D. Jung, 
Appl. Phys.B \textbf{93}, 349 (2008);
H.-C. Wu, 
J. Meyer-ter-Vehn, J. Fernandez, B. M. Hegelich, 
Phys. Rev. Lett. \textbf{104}, 234801 (2010); 
S. S. Bulanov, 
A. Maksimchuk, K. Krushelnick, K. I. Popov, V. Y.
Bychenkov, and W. Rozmus, 
Phys. Lett. A \textbf{374}, 476 (2010);
M. Wen {\it et al.}, Appl. Phys. Lett. {\bf 101}, 021102 (2012); 
H.-C. Wu and J. Meyer-ter-Vehn, Nat. Photonics {\bf 6}, 304 (2012);
J. K. Koga, 
S. V. Bulanov, T. Zh. Esirkepov, A. S. Pirozhkov, and M. Kando, 
Phys. Rev. E \textbf{86}, 053823 (2012).

\bibitem{Kulagin2013}  V. V. Kulagin, V. N. Kornienko, V. A. Cherepenin, 
H. Suk, Quantum Electronics {\bf 43}, 443 (2013).

\bibitem{Kando2007}M. Kando, Y. Fukuda, A. S. Pirozhkov, J. Ma, I.
Daito, L.-M. Chen, T. Zh. Esirkepov, K. Ogura,
T. Homma, Y. Hayashi, H. Kotaki, A. Sagisaka,
M. Mori, J. K. Koga, H. Daido, S. V. Bulanov,
T. Kimura, Y. Kato, and T. Tajima, Phys. Rev.
Lett. {\bf 99}, 135001 (2007); 
A. S. Pirozhkov, J. Ma,
M. Kando, T. Zh. Esirkepov, Y. Fukuda, L.-M.
Chen, I. Daito, K. Ogura, T. Homma, Y. Hayashi,
H. Kotaki, A. Sagisaka, M. Mori, J. K. Koga, T.
Kawachi, H. Daido, S. V. Bulanov, T. Kimura,
Y. Kato, and T. Tajima, Plasma Phys. {\bf 14}, 123106
(2007); 
M. Kando, A. S. Pirozhkov, K. Kawase, T.
Zh. Esirkepov, Y. Fukuda, H. Kiriyama, H. Okada,
I. Daito, T. Kameshima, Y. Hayashi, H. Kotaki,
M. Mori, J. K. Koga, H. Daido, A. Ya. Faenov,
T. Pikuz, J. Ma, L.-M. Chen, E. N. Ragozin, T.
Kawachi, Y. Kato, T. Tajima, and S. V. Bulanov,
Phys. Rev. Lett. {\bf 103}, 235003 (2009).

\bibitem{Kagami} T. Zh. Esirkepov, S. V. Bulanov, M. Kando, A. S. Pirozhkov, 
and A. G. Zhidkov, Phys. Rev. Lett. {\bf 103}, 025002 (2009).

\bibitem{BrII2012} A. V. Panchenko, T. Zh. Esirkepov, A. S.
Pirozhkov, M. Kando, F. F. Kamenets, and S. V.
Bulanov, Phys. Rev. E {\bf 78}, 056402 (2008).
S. V. Bulanov, T. Zh. Esirkepov, M. Kando, J. K.
Koga, A. S. Pirozhkov, T. Nakamura, S. S. Bulanov,
C. B. Schroeder, E. Esarey, F. Califano, and
F. Pegoraro, Phys. Plasmas {\bf 19}, 113102(2012); Phys. Plasmas {\bf 19}, 113103 (2012).

\bibitem{Shaping} S. V. Bulanov, T. Zh. Esirkepov, N. M. Naumova, F. Pegoraro, I. V. Pogorelsky, A. M. Pukhov, 
IEEE Trans. Plasma Sci. {\bf 24}, 393 (1996).

\bibitem{Reed2009} S. A. Reed, T. Matsuoka, S. S. Bulanov, M. Tampo, 
V. Chvykov, G. Kalintchenko, P. Rousseau, V. Yanovsky, R. Kodama, 
D. W. Litzenberg, K. Krushelnick, and A. Maksimchuk, Appl. Phys. Lett. {\bf 94}, 201117 (2009). 

\bibitem{Foil3} D. Kiefer, M. Yeung, T. Dzelzainis, P. S. Foster, S. G. Rykovanov, 
C. LS. Lewis, R.S. Marjoribanks, H. Ruhl, D. Habs, J. Schreiber, M. Zepf,
 and B. Dromey, Nat. Comm. \textbf{4}, 1763 (2013).

\bibitem{transparency} S. Palaniyappan, B. M. Hegelich, H.-C. Wu, D. Jung, D. C.
Gautier, L. Yin, B. J. Albright, R. P. Johnson, T. Shimada,
S. Letzring, D. T. Offermann, J. Ren, C. Huang, R. Horlein,
B. Dromey, J. C. Fernandez, and R. C. Shah, Nature Phys. {\bf 8},
763 (2012).

\bibitem{Hur2012} M. S. Hur, Y.-K. Kim, V. V. Kulagin, I. Nam, and H. Suk, Phys. Plasmas {\bf 19}, 073114 (2012).

\bibitem{Bulanov1975} S. V. Bulanov, 
Radiophys. Quantum Electronics {\bf 18}, 1511 (1975).

\bibitem{Bratman1995} V. I. Bratman and S. V. Samsonov, Phys. Lett. A {\bf 206}, 377 (1995).

\bibitem{LADvsLL} S. V. Bulanov, T. Zh. Esirkepov, M. Kando, J. Koga, 
and S. S. Bulanov, Phys. Rev E {\bf 84}, 056605 (2011).

\bibitem{Feynman1966} R. P. Feynman, R. P. Leiton, and M. Sands, 
{\it The Feynman Lectures on Physics, Vol. 2} (Addison-Wesley, Reading, MA, 1966) Section 18-4.

\bibitem{LandauLifshitzVol2}L. D.~Landau and E. M.~Lifshitz, {\it The Classical Theory of Fields}
(Pergamon, Oxford, 1975).

\bibitem{Bourdier} A. Bourdier, Phys. Fluids {\bf 26}, 1804 (1983).

\bibitem{PGARB} P. Gibbon and A. R. Bell, Phys. Rev. Lett. {\bf 68}, 1535 (1992).

\bibitem{Dawson62} J. Dawson, Phys. Fluids {\bf 5}, 445 (1962).


\bibitem{Einstein}A. Einstein, Ann. Phys. (Leipzig) \textbf{17}, 891 (1905).

\bibitem{Lichters94}R.~Lichters, J. Meyer-ter-Vehn, and  A. M. Pukhov, Phys. Plasmas {\bf 3}, 3425 (1996).

\bibitem{Hartemann2002} F. V. Hartemann, {\it High-Field Electrodynamics} (CRC Press, Boca Raton, FL, 2002).

\bibitem{LLCTF} L. D. Landau and E. M. Lifshitz, {\it The Classical Theory of Field}, (Pergamon, Oxford, 1975).

\bibitem{Lawson} J. D. Lawson, IEEE Trans. Nucl. Sci. {\bf NS-26}, 4217 (1979).

\bibitem{Woodward} P. M. Woodward, J. IEEE {\bf 93} Part III A, 1554 (1947).

\bibitem{Lobet} M. Lobet, M. Kando, J. K. Koga, T. Zh. Esirkepov, T. Nakamura, A. S. Pirozhkov, and S. V. Bulanov, 
Phys. Lett. A {\bf 377}, 1114 (2013).

\bibitem{RT-PB} F. Pegoraro and S.~V. Bulanov, 
Phys. Rev. Lett. {\bf 99}, 065002 (2007).

\bibitem{kagami} T. Zh. Esirkepov, S. V. Bulanov, M. Kando, A. S. Pirozhkov, A. G.  Zhidkov, 
Phys. Rev. Lett. {\bf 103}, 025002 (2009).

\bibitem{SKAR}	S. Kar, M. Borghesi, S.V. Bulanov, A. Macchi, M. H. Key, T. V. Liseykina, A. J. Mackinnon, 
P. K. Patel, L. Romagnani, A. Schiavi, and O. Willi, Phys. Rev. Lett. {\bf 100}, 225004 (2008); 
S. Kar,  K. F. Kakolee, B. Qiao, A. Macchi, M. Cerchez,  D. Doria, 
M. Geissler, P. McKenna, D. Neely, J. Osterholz, 
R. Prasad, K. Quinn, B. Ramakrishna, G. Sarri, O. Willi, X. Y. Yuan, 
M. Zepf, and M. Borghesi, Phys. Rev. Lett. {\bf 109}, 185006 (2012); 
F. Dollar, C. Zulick, A. G. R. Thomas, V. Chvykov, J. Davis, G. Kalinchenko, 
T. Matsuoka, C. McGuffey, G. M. Petrov, L. Willingale, V. Yanovsky, A. Maksimchuk, and K. Krushelnick, 
Phys. Rev. Lett. \textbf{108}, 175005 (2012); 
I. J. Kim, K. H. Pae, C. M. Kim, H. T. Kim, J. H. Sung, S. K. Lee,  T. J. Yu, I. W. Choi, C.-L. 
Lee, C. H. Nam, P. V. Nickles, T. M. Jeong, J. Lee, Phys. Rev. Lett. {\bf 111}, 165003 (2013).
 
 \bibitem{QEDint} C. P. Ridgers, C. S. Brady, R. Duclous, J. G. Kirk, K. Bennett, 
 T. D. Arber, A. P. L. Robinson, A. R. Bell,
Phys. Rev. Lett. {\bf 108}, 165006 (2012); A. G. R. Thomas, C. P. Ridgers, S. S. Bulanov, B. J. Griffin, and S. P. D. Mangles, Phys. Rev. X \textbf{2}, 041004 (2012);
 S. V. Bulanov, T. Zh. Esirkepov, M. Kando, J. K. Koga, T. Nakamura, 
 S. S. Bulanov, A. G. Zhidkov, Y. Kato, G. Korn, 
 Proceed. SPIE-2013 Int. Conf.,  SPIE Optics + Optoelectronics, pp. 878015-878015-15 (2013);
 S. S. Bulanov, C. B. Schroeder, E. Esarey, W. P. Leemans,
Physical Review A {\bf 87}, 062110 (2013).


\end{thebibliography}
\end{document}